\pdfoutput=1
\documentclass[fleqn,usenatbib]{mnras}

\usepackage{newtxtext,newtxmath}
\usepackage[T1]{fontenc}
\DeclareRobustCommand{\VAN}[3]{#2}
\let\VANthebibliography\thebibliography
\def\thebibliography{\DeclareRobustCommand{\VAN}[3]{##3}\VANthebibliography}
\usepackage[normalem]{ulem}

\usepackage{graphicx}	% Including figure files
\title{Constraining black hole feedback in galaxy clusters from X-ray power spectra}

\author[A. Heinrich, Y.-H. Chen, S. Heinz et al.]{
Andrew M. Heinrich,$^{1}$\thanks{E-mail: amheinrich@wisc.edu}
Yi-Hao Chen,$^{1}$
Sebastian Heinz,$^{1}$
Irina Zhuravleva,$^{2}$
Eugene Churazov$^{3}$
\\
$^{1}$Department of Astronomy, University of Wisconsin Madison, 475 N. Charter Street, Madison, WI 53706\\
$^{2}$Department of Astronomy and Astrophysics, University of Chicago, 5640 South Ellis Avenue, Chicago, IL 60637\\
$^{3}$Max Planck Institute for Astrophysics, Karl-Schwwarzschild-Strasse 1, D-85478 Garching, Germany}

\date{Accepted 2021 May 24. Received 2021 May 03; in original form 2020 December 17}

\pubyear{2021}

\begin{document}
\label{firstpage}
\pagerange{\pageref{firstpage}--\pageref{lastpage}}
\maketitle

\begin{abstract}
Jets launched by the supermassive black holes in the centers of cool-core clusters are the most likely heat source to solve the cooling flow problem.
One way for this heating to occur is through generation of a turbulent cascade by jet-inflated bubbles. 
Measurements of the X-ray intensity power spectra show evidence of this cascade in different regions of the cluster, constraining the role of driving mechanisms. 
We analyze feedback simulations of the Perseus cluster to constrain the effect of the jet activity on the intensity fluctuations and kinematics of the cluster atmosphere. 
We find that, within the inner 60 kiloparsecs, the power spectra of the predicted surface brightness fluctuations are broadly consistent with those measured by Chandra and that even a single episode of jet activity can generate a long-lasting imprint on the intensity fluctuations in the innermost region of the cluster.
AGN-driven motions within the same region approach the values reported by Hitomi during and right after the AGN episode.
However, the line-of-sight velocity dispersion excited by the jet in simulations underpredicts the Hitomi measurement.
This indicates that driving a volume-filling sustained level of turbulence requires several episodes of jet activity, and/or additional processes drive turbulence outside the 60-kpc sphere.
This also suggests that sharp edges of the bubbles in the innermost region of the cluster contribute substantially to the intensity of fluctuations, consistent with the Perseus observations in the inner 30-kpc region.
We discuss new diagnostics to decompose annular power spectra to constrain past episodes of jet activity.
\end{abstract}

\begin{keywords}
galaxies: clusters -- methods: analytical -- clusters: intracluster medium -- galaxies: jets -- MHD
\end{keywords}

%%%%%%%%%%%%%%%%%%%%%%%%%%%%%%%%%%%%%%%%%%%%%%%%%%

%%%%%%%%%%%%%%%%% BODY OF PAPER %%%%%%%%%%%%%%%%%%

\section{Introduction}

    \begin{figure}
        \includegraphics[width=\columnwidth]{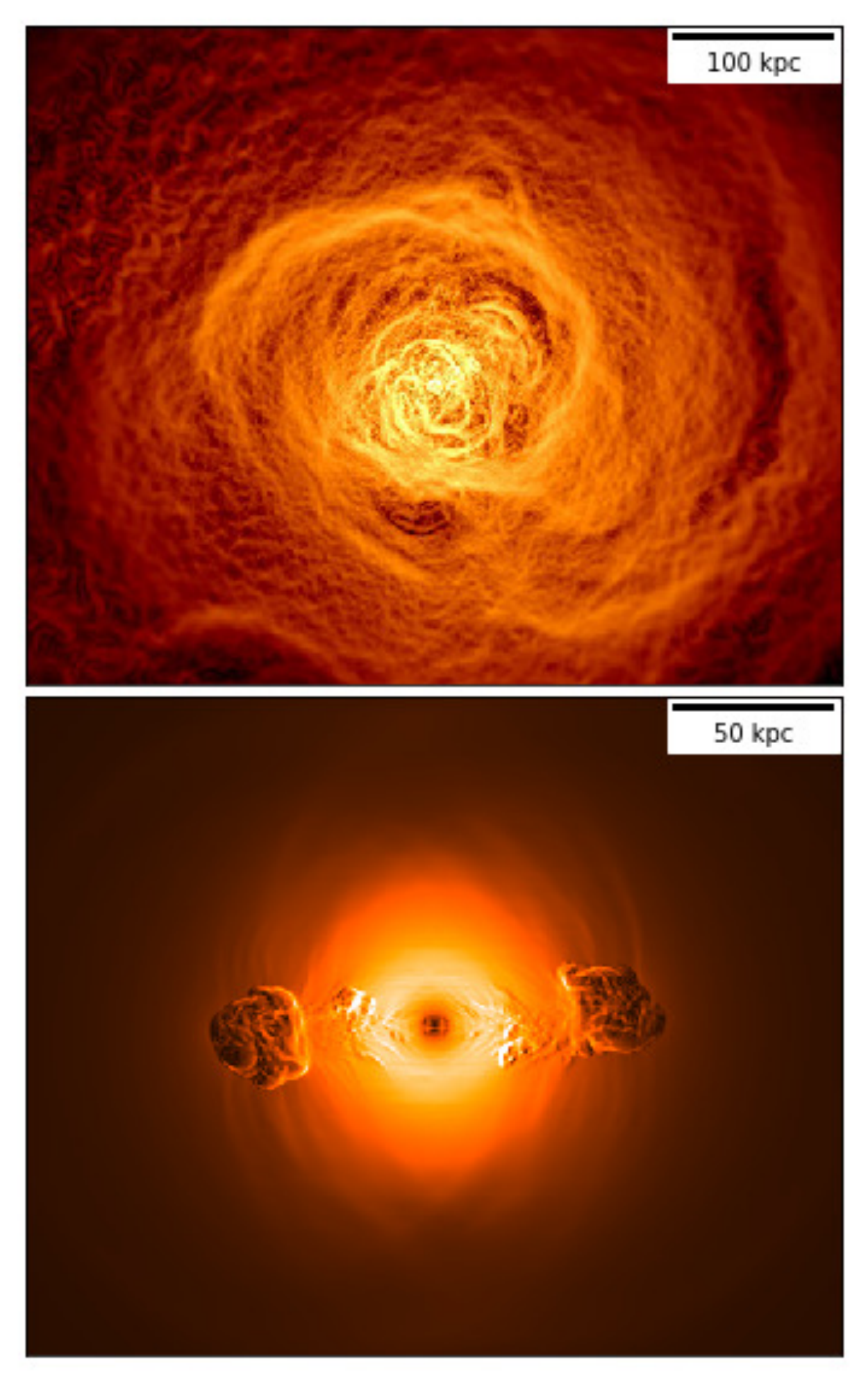}
            \caption{Top:Gaussian Gradient Magnitude filtered image of the Perseus Cluster from \citet{Sanders2016}. Bottom: A similarly treated image of our simulation at 50 Myr, through a GGM filter, for context. The \citet{Sanders2016} image uses GGM filters on a variety of scales, with the center of the image using smaller $\sigma$ values than the edges. The simulation image is treated with a single GGM filter with $\sigma=1$ pixel ($0.5$ kpc)}
        \label{fig:ggm}
    \end{figure} 
 
    Chief among the largest structures in the universe are galaxy clusters, huge collections of gravitationally bound galaxies surrounded by X-ray emitting gas, held together by the gravity of their host dark matter halo. 
    The cluster atmosphere should be losing energy and cooling due to this X-ray emission, with cooling times substantially shorter than the age of the cluster (see \citealt{Fabian1994} for review). 
    The cooled gas should then sink to the center of the cluster and coalesce at a rate upward of 100 solar masses per year in the largest clusters \citep[e.g.][]{Fabian1994}, triggering star formation at similar levels. 
    However, observations of star formation only show a small fraction of this rate and X-ray spectra indicate a lack of cold gas \citep[e.g.][]{Peterson2001,Peterson2002}, leading to the conclusion that some form of heating is preventing the formation of a cooling flow. 
    
    Several sources of energy for powering clusters have been suggested. 
    Feedback from jets launched by central brightest galaxy's supermassive black hole is the most promising candidate \citep[e.g.][]{burns:90,Churazov2000,Fabian2000, mcnamara:07}, but other sources of energy have been considered as supplemental or alternative sources of energy input that could counteract the cooling, including merger-driven turbulence and sloshing \citep[e.g.][]{motl:04,rossetti:10,parrish:10,su:16,su:17}, thermal conduction \citep[e.g.][]{medvedev:01,Yang2016,Chen2019}, and cosmic ray streaming \citep[e.g.][]{sharma:10,ruszkowski:17,ehlert:18}. 
    The cooling flow problem has been a central issue in galaxy cluster astrophysics for nearly 30 years. 

    The archetypal galaxy cluster for these observations is the Perseus cluster, the brightest cluster in the X-ray band. 
    It has some of the deepest observations from the {\em Chandra X-ray Observatory} (Figure \ref{fig:ggm}), {\em XMM Newton}, and {\em Suzaku}, \citep{Fabian2003, Bulbul2014, Edge1990} as well as the only measurement taken from the {\em Hitomi} satellite \citep{Aharonian2016,aharonian:18b}.
    These observations reveal a variety of structures all of which are theorized to come from different processes. 
    The inner core of the cluster is dominated by the jet-inflated radio cavities from the active galactic nucleus (AGN) in NGC1275 \citep{Birzan2004}, while the outer core’s structure can be attributed to sloshing from an ongoing galaxy merger \citep{Churazov2003, Fabian2011, Walker2018, zuhone2011}.

    The level of turbulence in Perseus has been constrained by several datasets, most notably the {\em Hitomi} satellite, which measured the small-scale velocity dispersion in Perseus to be  164 +/- 10 km/s \citep{Aharonian2016}. 
    Prior to this measurement, there was a large effort in the cluster community to predict what {\em Hitomi} would see. 
    The seminal method for this was to use amplitudes of X-ray image fluctuations from {\em Chandra} to quantify turbulence in the atmosphere via estimated power spectra.

    By first removing large scale density gradients using a beta model fit to the cluster surface brightness, \citet{Zhuravleva2015} calculate the density fluctuation spectra for different annuli of the Perseus {\em Chandra} images. 
    Assuming a linear relation with the velocity spectrum, they found turbulence in agreement with the {\em Hitomi} measurement, between 90 and 140 km/s on scales \textasciitilde 30-40 kpc. 
    %IZ%The largest fluctuations are measured in the innermost 30 kpc, due to bubbles inflated by the AGN.
    This level of turbulence is sufficient to balance observed radiative cooling and explain the mass deposition gap \citep{Zhuravleva2014}.

    Simulations of galaxy clusters have become integral to isolating different sources of turbulence in galaxy clusters. 
    While the power spectrum technique for X-ray bright clusters was first implemented as a predictive method, it has since been used as an analytical tool to measure the contribution of different simulated processes to the overall turbulence of the cluster. 
    It has been shown that brightness fluctuations outside of the inner 60 kpc of Perseus the cluster atmosphere can be entirely explained by gas sloshing. 
    Using the sloshing simulations of \citet{ZuHone2018}, \citet{Walker2018} measured the power spectra in different annuli in a galaxy cluster perturbed by a merger like Perseus, comparing the power spectra from each. 
    We adopt this same method and apply it to simulations that include feedback by AGN jets in the Perseus Cluster in order to further constrain the source of turbulent heating of the ICM. 
    We implement a $\Delta$-variance method for power spectrum calculation in annuli of varying radius and sample multiple jet ages to construct an approximate spectrum composed of multiple jet episodes.
    
    The paper is organized as follows: Section \ref{sec:methods} describes the methodology of the simulations used, the power spectra derivation, and modelling of multiple jet episodes. We present results in Section \ref{sec:results}, and discuss our findings, with comparison to velocity dispersion measurements, in Section \ref{sec:discussion}. Section \ref{sec:conclusions} summarized the conclusions of the paper.
    
\section{Methods}
    \label{sec:methods}
    \subsection{Numerical Simulation}
    \label{sec:simul} 
    \begin{figure}
        \centering
        \includegraphics[width=\columnwidth]{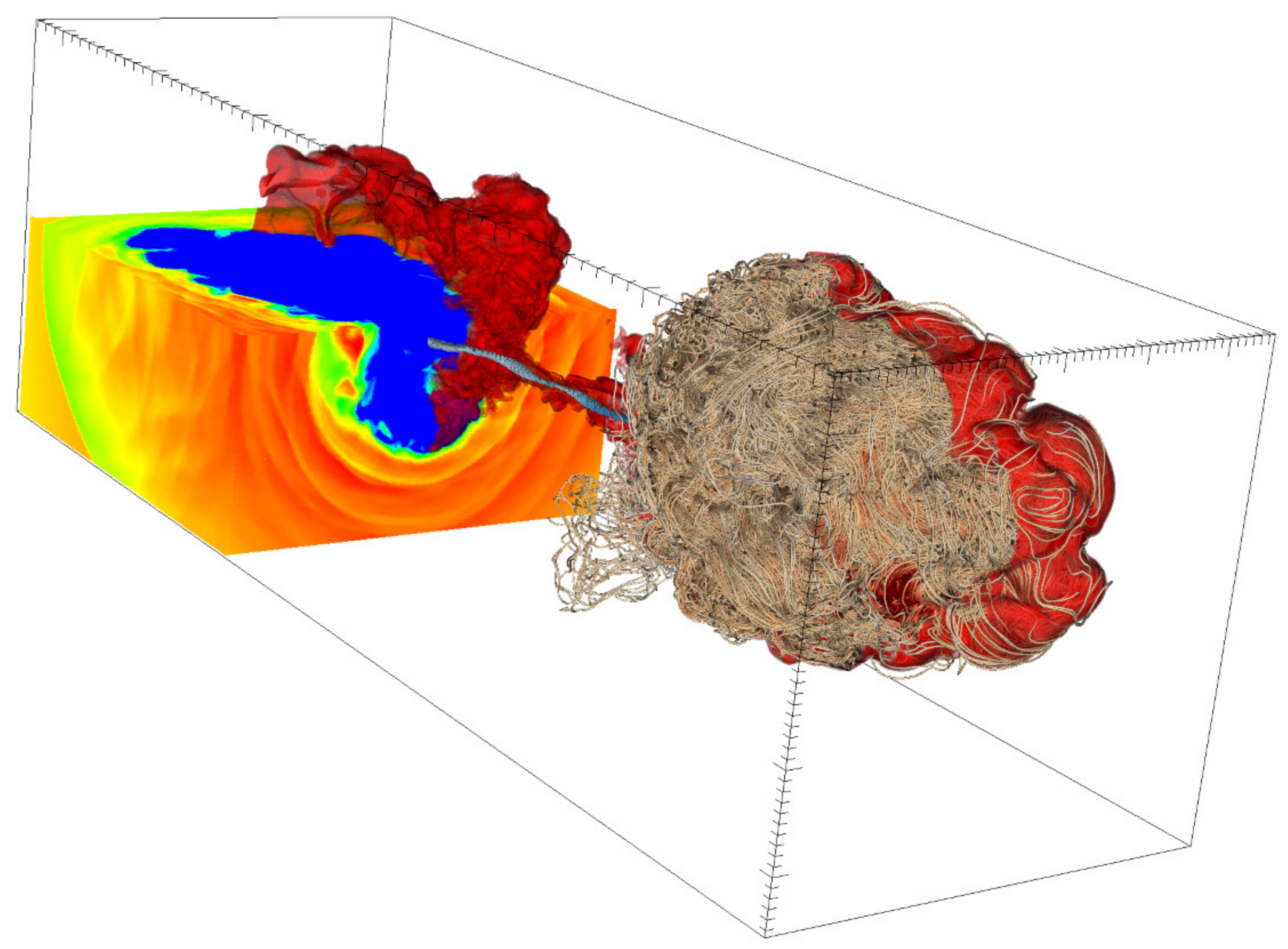}
            \caption{A three-dimensional rendering of the simulation; magnetic field lines are rendered as tubes, the red surfaces display iso-density contours, highlighting the radio lobes/cavities, and the colored slice plots in the lower rear quadrant indicate temperature variations.}
        \label{fig:jetsim}
    \end{figure}
    
    We use magnetohydrodynamic simulations with an adaptive mesh refinement (AMR) grid. 
    A more in-depth discussion of the numerical set up is included in \citet{Chen2019}, where the simulations used in this paper were first described.

    The simulations were performed using the publicly available FLASH code \citep{Dubey2009}, modeling an AGN jet embedded in a spherically symmetric atmosphere of monatomic ideal gas, distributed throughout the simulation box (0.5 Mpc by 0.5 Mpc by 1 Mpc). 
    The cluster atmosphere follows a $\beta$-model in density, calibrated to  match observations of the Perseus Cluster atmosphere and adopts the temperature profile from \citet{Zhuravleva2015}:
    \begin{equation}
      \rho(r)=\rho_{0}\left[1 + \left(\frac{r}{r_{\rm c}}\right)^2\right]^{-3\beta/2}
    \end{equation}
    \begin{equation}
      T(r)=T_{\rm out}\left[1 + \left(\frac{r}{r_{\rm c,t}}\right)^3\right]\left[\frac{T_{\rm out}}{T_{\rm core}} + \left(\frac{r}{r_{\rm c,T}}\right)^3\right]^{-1}
    \end{equation}
    with $\rho_{0}=9.6\times 10^{-26}\,{\rm g\,cm^{-3}}$, $r_{\rm c}=26\,{\rm kpc}$, $\beta=0.53$, $T_{\rm core}=3.0\,{\rm keV}$, $T_{\rm out}=6.4\,{\rm keV}$, and $r_{\rm c,T}=60\,{\rm kpc}$.
    
    The base of the jet is modeled as a nozzle at the center of the cluster \citep{heinz:06,Chen2019} which injects energy, mass, momentum, and magnetic flux into the atmosphere in both directions along the z-axis, with a small wobble added to account for the precession of the spin axis of the black hole and dynamical instabilities in the jet below the resolution scale, as an effective implementation of the "dentist drill"  effect first postulated by \citet{Scheuer1982}. 
    The simulation models the magnetic field in the jet, injecting magnetic field in equipartition with the particle pressure and a helicity of $B_{\phi}/B_{\rm z} = 1$, however, the ICM is nonmagnetized. 
    While the ICM in clusters is magnetized, the magnetic field is likely dynamically weak. 
    Modeling jet propagation into magnetized ICM is beyond the scope of the current paper and subject to future work.
    
    The structure of the simulation can be seen in a snapshot rendering in Figure \ref{fig:jetsim}. 
    The jet fires for 10 Myr at a power of $P_{\rm jet}=10^{45}\,{\rm ergs/s}$ before being shut off and allowed to propagate through the cluster for an additional 500 Myr. 
    The AMR resolution of the cluster gas changes throughout the simulation, based on the propagation of the jet, increasing refinement where the most activity occurs and decreasing resolution in low-activity areas to maximize computational efficiency \citep{Chen2019}. 
    The original simulation was designed to study the magnetic properties of the jets and as such required a high resolution of 30 pc to fully resolve the jets. After the jet is shut off at 10 Myr, the maximum resolution is reduced to 120 pc in the inner cluster for computational efficiency. In either case, the numerical resolution significantly exceeds the imaging resolution of {\em Chandra} at the distance of the Perseus cluster.

 \subsection{Synthetic X-ray Images}
    The simulations are largely analyzed using the yt package \citep{Turk2010} in conjunction with NumPy \citep{Harris2020} and Astropy \citep{astropy:2018}. 
    We produce a facsimile of observational data from our simulations via a projection of the X-ray emissivity along the coordinate axes of the grid. 
    The X-ray emissivity from 0.5 to 3.5 keV is calculated in each cell in the domain via the APEC code \citep{Smith2001}, assuming a uniform metallicity of half the solar value, using abundances from \citet{Anders1989}. 
    The full domain of the simulation is then integrated along a coordinate axis, creating a 0.5 Mpc by 0.5 Mpc image if the axis is along the jet, or 0.5 Mpc by 1 Mpc image if the projection is perpendicular to the jet-axis. 
    
\begin{figure}
        \includegraphics[width=\columnwidth]{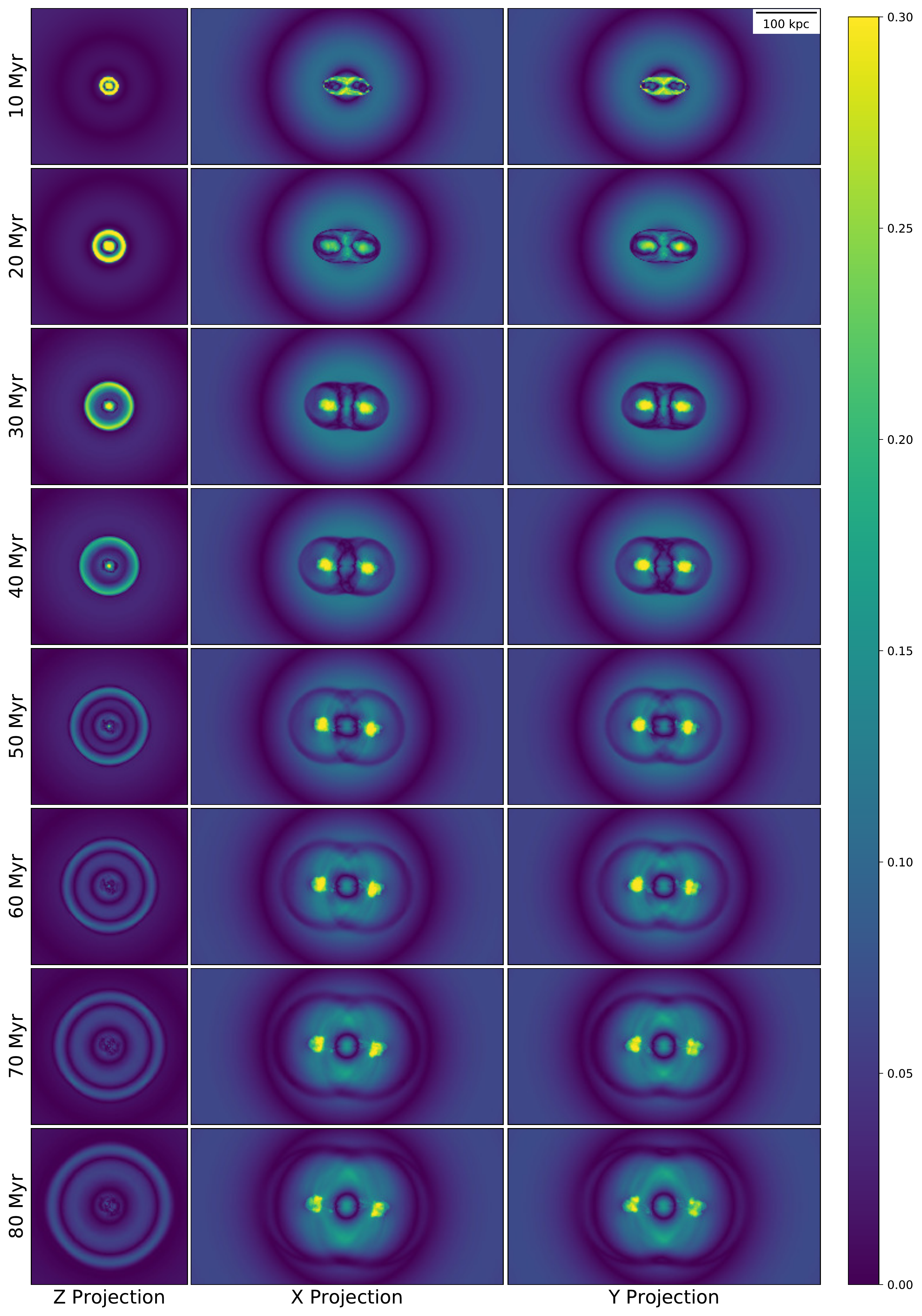}
        \caption{ Absolute fractional intensity deviations from azimuthal best-fit $\beta$-model in the 0.5-3.5 keV band for eight different simulation times shown at three viewing angles. $I=\frac{\lvert P-\beta \rvert}{\beta}$, where I is the normalized image seen in this figure, P is the projected X-ray surface brightness and $\beta$ is the azimuthally averaged $\beta$-model.}
        \label{fig:projfig}
      \end{figure}

    To reproduce the normalization of the images by a beta model used in \citet{Zhuravleva2015}, which we consider as the benchmark in this work, these images are then azimuthally averaged and fitted with a $\beta$-model: 	
	\begin{equation}
	    S= S_0\left[ 1+ \left( \frac{r}{r_c}\right)^2\right]^{-3 \beta +0.5}
	    \label{eq:betamod}
	\end{equation}
    with the best fit determined via a Monte Carlo markov chain (MCMC) sampler, which finds the most likely parameters $S_0$,  $\beta$, and the core radius $r_c$ for each individual image. 
    This unique $\beta$-model is then used to normalize the images, removing the large scale structure of the image and leaving an image of gas perturbation in the cluster. 
    Even though the simulation is set up with a beta model as the density profile, it is important to re-normalize the synthetic images with the best-fit beta model because the jet action affects the overall cluster intensity profile. 
    As the observational analysis by \citet{Zhuravleva2015} and \citet{Churazov2012} normalizes the images by a $\beta$-model fit to the currently observed cluster profile, our analysis aims to employ a similar methodology to each frame. 
    The final product of this process is seen in Figure \ref{fig:projfig}, and fitted $\beta$-model parameters are described in Table \ref{tab:beta}.
    
    \begin{table}
     \caption{Fitted $\beta$-model parameters of the on-axis projections. The central surface brightness, core radius, and $\beta$ parameter are fairly consistent after the jet has turned on.}
     \label{tab:beta}
     \begin{tabular}{lccc}
        \hline
        Time & Surface Brightness & Core Radius & Beta Parameter\\
        Myr & $10^{-4}$ ergs/cm$^2$/s/Sr & kpc & \\
        & $S_0$ & $r_c$ & $\beta$\\
        \hline
         0 & 1.224 & 26.0 & 0.530\\
        20 & 2.16$\pm$0.04 & 31.1$\pm$0.5 & 0.556$\pm$0.002\\
        40 & 1.91$\pm$0.02 & 33.7$\pm$0.3 & 0.560$\pm$0.054\\
        60 & 2.06$\pm$0.01 & 31.8$\pm$0.1 & 0.555$\pm$0.001\\
        80 & 2.20$\pm$0.01 & 30.3$\pm$0.1 & 0.551$\pm$0.001\\
        \hline
     \end{tabular}
    \end{table}
    
	Due to the AMR grid changing over the course of the simulation and throughout the domain, the normalization process reveals strong grid-like artifacts at large radii in the X-ray maps (where the AMR resolution is lower, enhancing differences in the discretized grid and the smooth beta model).
	Such artifacts would interfere with the power spectral measurement we aim to perform. 
	These artefacts can be removed by applying a smoothed covering grid (as implemented in the yt package), which creates a 3D grid structure at a fixed resolution, and performs a first order interpolation within coarse cells in order to fit the entire simulation into one resolution. 
	This ensures that the AMR grid does not interfere with the calculation of the power spectrum, within the limits of the linear interpolation scheme used by the algorithm \citep{Turk2010}.
	
    \subsection{Power Spectrum Calculation}
	In order to find a quantitative measure of the surface brightness fluctuations (and thus the level of turbulence), the X-ray maps of the cluster are divided into equally spaced annuli by masking the images. 
	Each annulus has a set width of \textasciitilde 30 kpc, following \citet{Zhuravleva2015}, corresponding to an angle of 1.5 arcminutes on the sky. 
	This allows us to quantify the level of surface brightness fluctuations in different regions of the cluster, which will be affected in different ways by various sources of driving. 
	Dividing the image like this means that we cannot calculate a power spectrum in the normal method, by averaging the two-dimensional Fourier transform of the image. Using the standard fast Fourier transform (FFT) method would result in a power spectrum dominated by the mask's edge effects. 
	Instead, we employ the $\Delta$-variance method \citep{Arevalo2012, Stutzki1998} for finding the power spectrum, employing a Mexican Hat filter to determine the power at each spatial frequency.
	We convolve the normalized image with a modified Mexican Hat filter (F):
	\begin{equation}
	F=\left( \frac{G_{\sigma 1}*I}{G_{\sigma 1}*M}-\frac{G_{\sigma 2}*I}{G_{\sigma 2}*M}\right) M
	\label{eq:mexican}
	\end{equation}
	where I is the image of the cluster, M is the annulus mask, and $G_{\sigma 1}$ and $G_{\sigma 2}$ are closely separated Gaussian filters with $\sigma_1=\sigma/ \sqrt{1+\epsilon}$ and $\sigma_2=\sigma \sqrt{1+\epsilon}$ ($\epsilon <<1$). 
	We can then integrate over the square of the filtered result (Equation \ref{eq:variance}) to obtain the Variance $V_k$, yielding the power spectrum $P(k)$.
	\begin{equation}
    \label{eq:variance}
	    V_k=\int (F*I)^2d^2x=P(k) \epsilon^2 \pi k^2  
	\end{equation}
	Estimating a three-dimensional power spectrum from this two-dimensional spectrum by relating them to a factor of the integrated square of the power spectrum along the line of sight (Equation \ref{eq:3d}), and converting to a density fluctuation spectrum $A_{3D}= \sqrt{4\pi P_{3D}k^3}$, allows us to compare the turbulence in each annulus at each time in the simulation to the X-ray brightness images from {\em Chandra} \citep{Zhuravleva2015}.
	\begin{equation}
	    \label{eq:3d}
	    \frac{P_{2D}(k)}{P_{3D}(k)} \approx 4\int |W(k_z)|^2 dk_z 
	\end{equation}
	We verified the fidelity of our implementation on filtered images injected with known power spectra.
	The variance method accurately reproduces the injected power spectrum and matches the power spectrum measured by traditional FFT when applied to an unmasked image. 
	Our analysis includes wavefactors corresponding to scales from 3 to 150 kpc on a logarithmic scale.

    \begin{figure*}
        \centering
        \includegraphics[width=\textwidth]{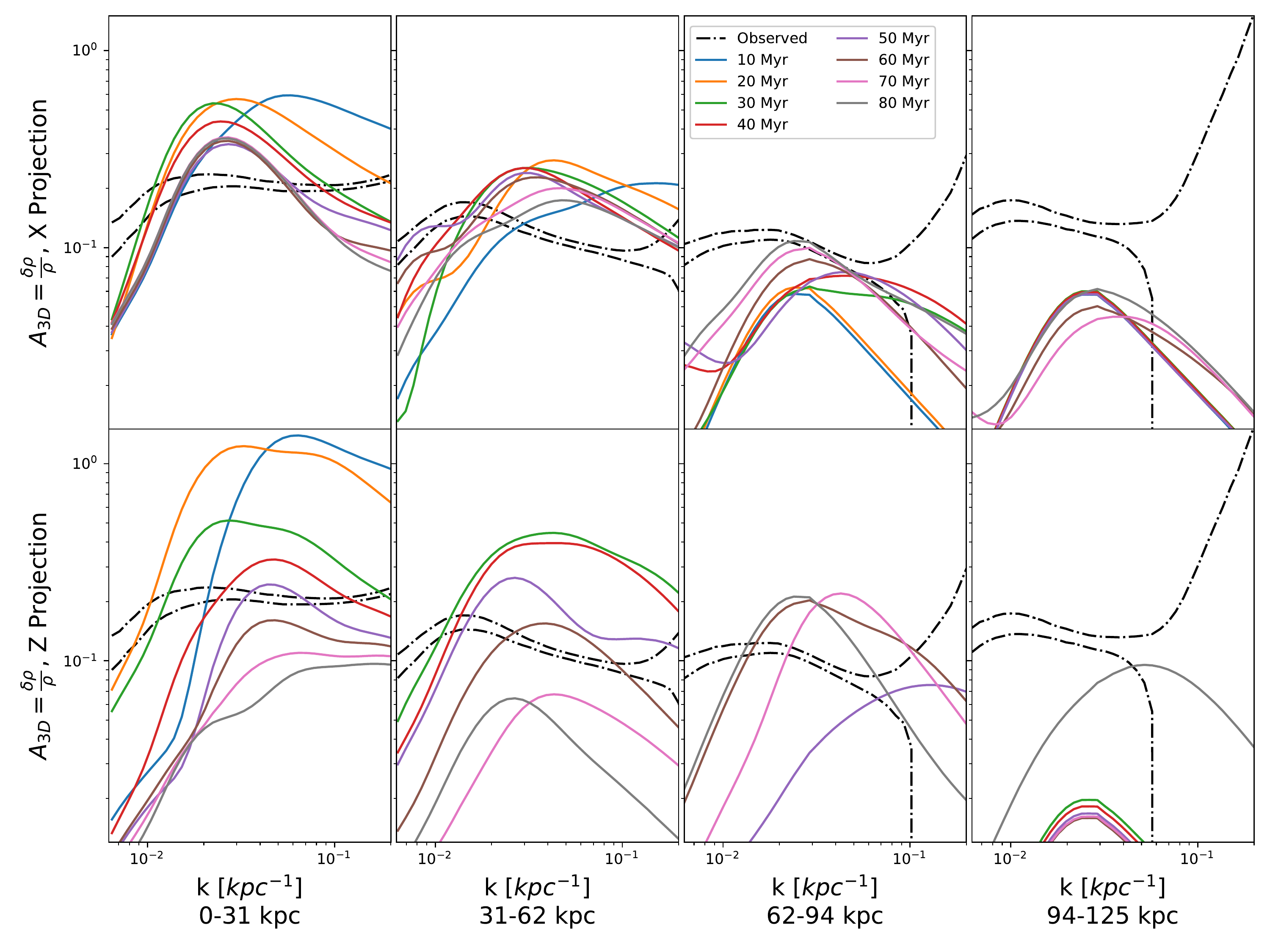}
        \caption{Spectra taken from the first four annuli in the x and z directions, overplotted on the observed PSD from \citet{Zhuravleva2015} at $1 \sigma$ uncertainty. Vertical axis is given as the three dimensional density fluctuations spectrum derived from the power spectrum of X-ray surface brightness fluctuations.}
        \label{fig:annspec}
    \end{figure*}
        
\subsection{Fitting the Data}
	The presence of at least two sets of cavities shows that the Perseus cluster has undergone a varied and complex history of jet activity that still affects the cluster atmosphere.  
	As such, any one snapshot of our simulation of a single 10 Myr jet event will not be able to account for the level of fluctuations seen in the observational data. 
	Because we are limited by computing time and cannot simulate infinite variations of jet power and duty cycles, we instead look to approximate a complex jet history by combining power spectra from different simulation times. 
	We sample the simulation every 10 Myr from the beginning of the simulation up to 80 Myr. 
	This gives us a wide basis that includes images of the cluster just after the jet has shut off as well as later times when the jet perturbations have had time to propagate through the cluster. 
	In order to find the best approximation of our single-event simulation to the Perseus cluster, which has undergone multiple AGN episodes, each under varied conditions, we attempt to model the power spectrum of a complex cluster by combining the power spectra of different simulation times. 
	If we approximate the power spectra as linear perturbations of the cluster atmosphere, we can find a linear combination of our basis that best fits the observed turbulence of the cluster atmosphere.

	To do this, we implement a Markov chain Monte Carlo (MCMC) algorithm \citep{Hogg2018} to find the best linear fit of our simulated power spectra to the observed spectra from the {\em Chandra} observation of the Perseus cluster. 
	The MCMC algorithm uses the emcee package \citep{Foreman-Mackey2012} to explore the likelihood space and sample many combinations of each weighting coefficient for the 8 basis spectra, measuring the likelihood function of each. 
	This algorithm gives us a large sample in parameter space of each time in the simulation, and its success is less dependent on initial conditions than a least-squares fit. 
	We can run this fitting algorithm on any combination of annuli in order to find where in the cluster the turbulence is most readily explained by the AGN jet. 
	The fitting process also ensures that the outer annuli are not overpredicted by the combination of jet episodes.

\section{Results}
    \label{sec:results}
    The power spectra derived from the AGN simulation show a wide range of fluctuations throughout the life cycle of the cluster. 
    In the earliest times, when the impact of the jet on the cluster has only reached the first or second annulus, the power spectral signature of the activity can reach levels equivalent to up to three times the measured {\em Chandra} surface brightness fluctuations of the cluster (Figure \ref{fig:annspec}). 
    Later in the simulation, after the jet has propagated for $>60$ Myr, the amplitude of the power spectrum in the outer annuli increases and approaches, but does not reach, the observed values in \citet{Zhuravleva2015}, while the power spectral amplitude in the inner 60 kpc decreases. 
    
    However, while the power in the outer annuli is below the observed values, for the projections perpendicular to the jet axis, the power in the inner annuli never decreases below the observations, reflecting the fact that the perturbations are driven by, and concentrated around, the central engine. 
    This suggests that the perturbations generated by jet-driven cavities alone cannot explain the observed fluctuations in clusters by themselves at all radii, while they can readily account for the observed power within the cool core of the Perseus cluster. 
    Thus, a combination of AGN activity and sloshing appears to be required to account for the observed power spectra in Perseus (see discussion in \S\ref{sec:sloshing}).
   
\begin{figure}
        \centering
        \includegraphics[width=\columnwidth]{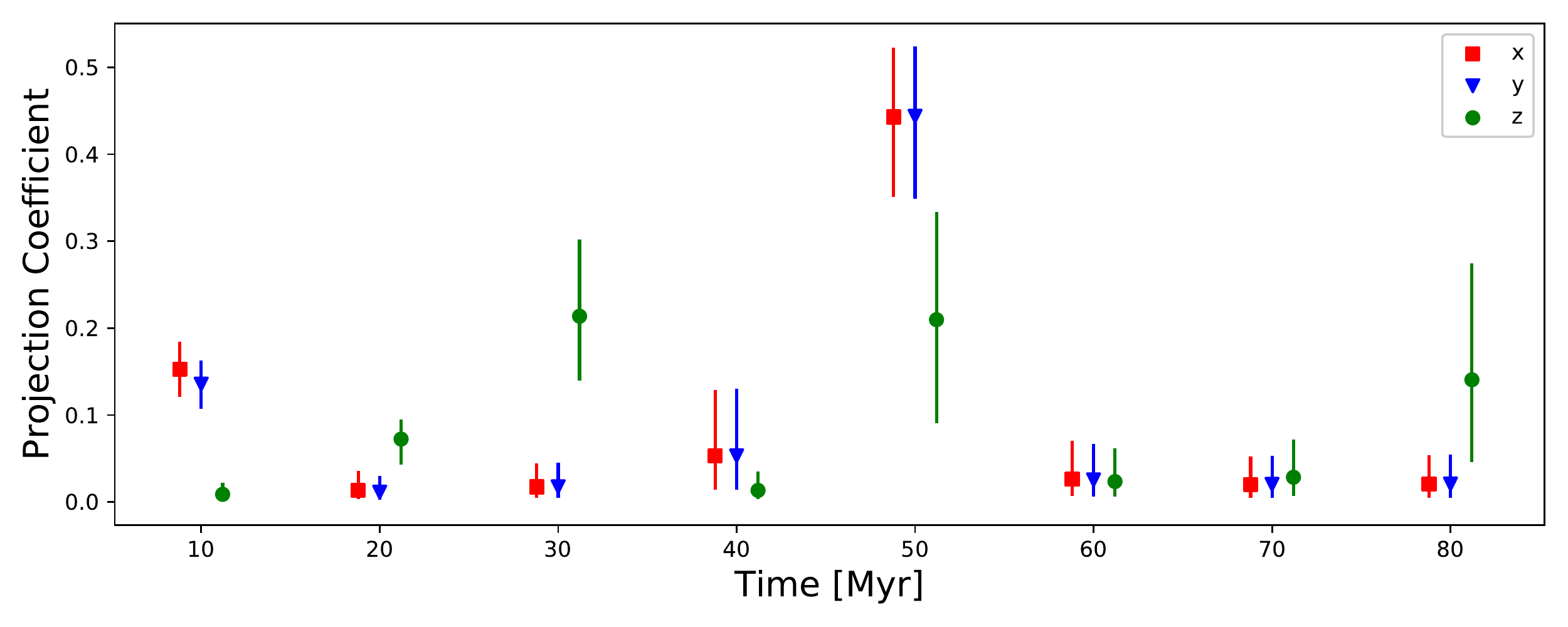}
        \caption{Weighted coefficients of sampled power spectra from three viewing angles. The fitted off-axis projection is dominated by the 10 Myr and 50 Myr jets, while the on-axis projection is made up of the 30, 50, and 80 Myr jets. /textcolor{red}{The sum of the weighted coefficients iss less than 1, indicating our simulation over-predicts the surface brightness fluctuations. See Section \ref{sec:spectraldensity} for more discussion.}}
        \label{fig:coeffs}
    \end{figure}

\begin{figure*}
        \centering
        \includegraphics[width=\textwidth]{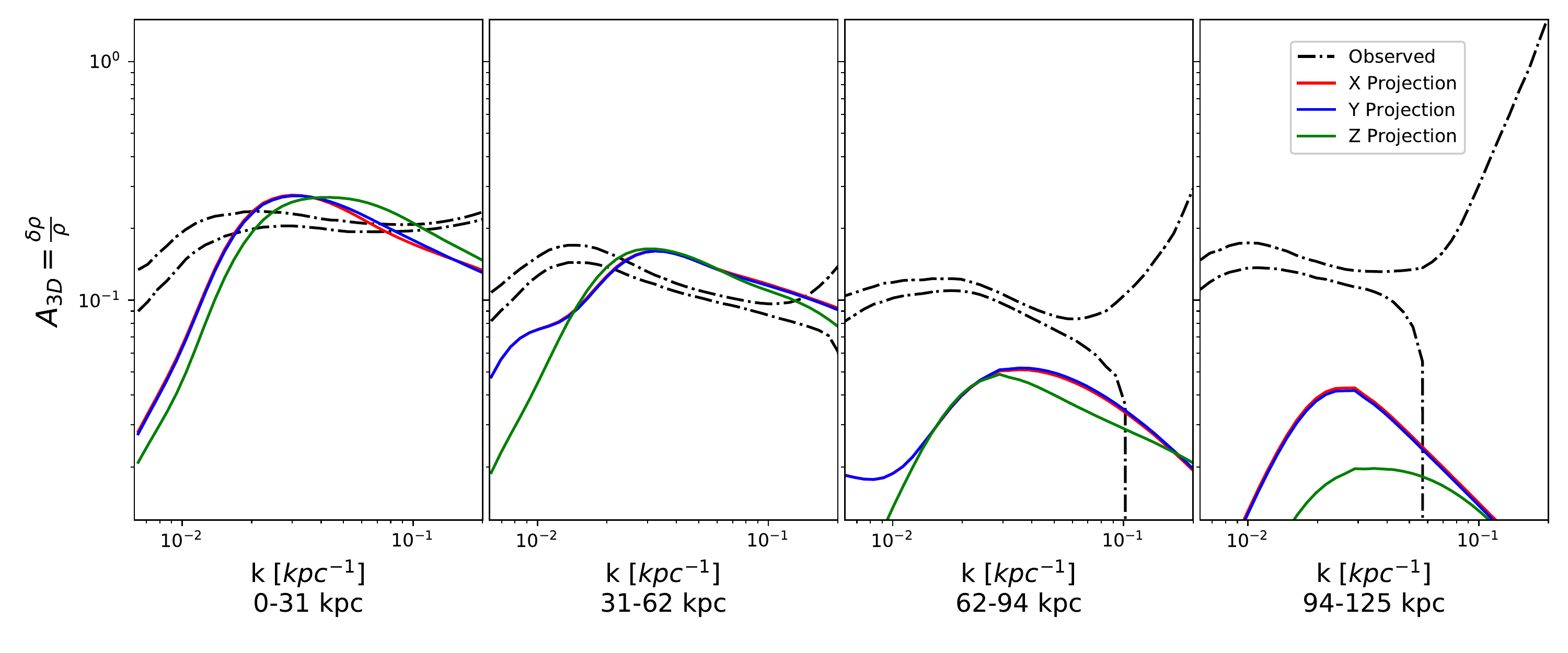}
        \caption{Combined weighted power spectra at three viewing angles overplotted on observed minimum and maximum PSD from \citet{Zhuravleva2014}}
        \label{fig:superspec}
    \end{figure*}

    In order to explore whether the power spectra measured in the inner cluster can be explained by past and/or current AGN activity, we use our MCMC fitting algorithm on the inner 60 kpc of the cluster and treat the {\em Chandra} power spectral data in the annuli larger than 60 kpc as upper limits, ensuring that the fitted spectra outside of the first two annuli are underpredicted by the fitting model. 
    This produces the best-fit weighting coefficients of the power spectra at each sampled time. 
    This best fit of the power spectra is dominated by the 10 and 50 Myr projections at the x and y viewing angles (line-of-sight perpendicular to mean jet axis), and the 30, 50, and 80 Myr projections when the jet is viewed in the z direction (along the jet axis), seen in Figure \ref{fig:coeffs}, while the fitted spectra itself is shown in Figure \ref{fig:superspec}
    
\begin{figure}
        \centering
            \includegraphics[width=\columnwidth]{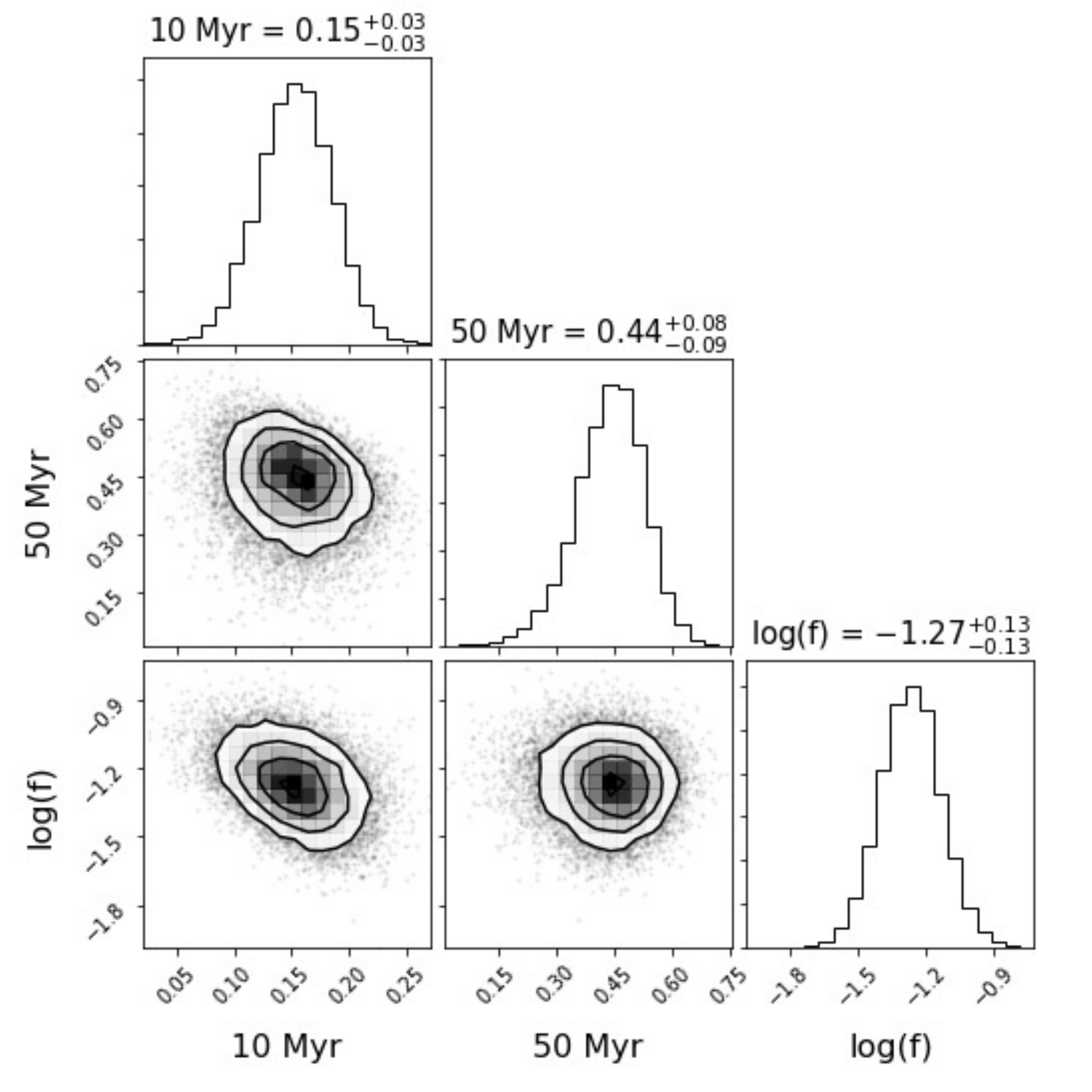}
            \includegraphics[width=\columnwidth]{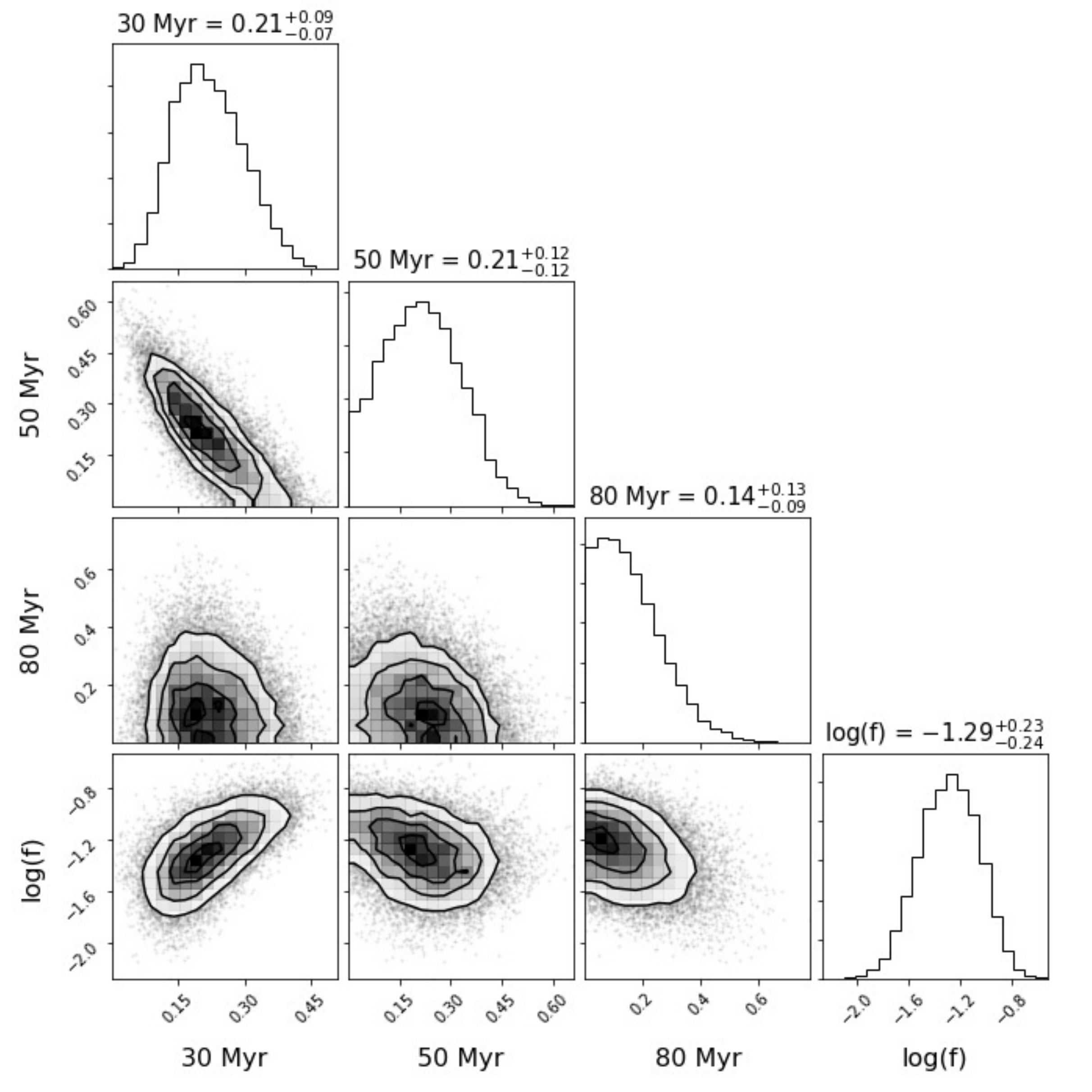}
        \caption{Top: on-axis projection. Bottom: off-axis projection. These corner plots show both the one and two dimensional projections of the probability distributions of the dominant jet ages in each linear combination. The two-dimensional projection shows the covariant dependencies between parameters.}
        \label{fig:corner}
      \end{figure}

    The corner plot in Fig.~\ref{fig:corner} shows each coefficient and its distribution of most likely values taken from the MCMC fit. 
    The MCMC algorithm samples \textasciitilde 3500 combinations of the fitted parameters in order to find the most likely combination of each. Each column corresponds to a sampled simulation time; e.g., the column marked 20 Myr is the contribution of the 20 million year-old projection compared to the other fitted coefficients. 
    In each column the relative linear contribution of power injected by a jet of the given age is compared to the rows corresponding to the power of different jet ages. 
    The contours in each cell demonstrate the covariant dependencies of the different parameters. 
    Each column is also end-capped with the total confidence histogram marginalized over the other parameters.
    
\section{Discussion}
    \label{sec:discussion}
    \subsection{The Power Spectral Density Profiles}
    \label{sec:spectraldensity}
    As is clear in Figure \ref{fig:annspec}, the evolution of the density fluctuation spectrum across the life of the simulation is not a trivial progression, especially when compared to the {\em Chandra} measurements. 
    The first 40 Myr of the simulation only show significant perturbations in the inner 60 kpc, at levels somewhat higher than, but comparable to the observed turbulence in the cluster \citep{Zhuravleva2015}.
    
    The spectra are also typically suppressed compared to the observed levels at low-k values, i.e., at larger spatial scales, leveling off after the peak between 0.02 and 0.05 $kpc^{-1}$, with the peak of the power spectral density (PSD) at slightly higher $k$ values than observed. 
    This may be due to the lack of any large scale asymmetry in the initial conditions of the simulation, compared to the large scale asymmetries observed in the X-ray maps of Perseus, even after the removal of asymmetries in the {\em Chandra} images, which would introduce power unrelated to the AGN at small $k$ values \citep[see dotted regions at small $k$ in][]{Zhuravleva2015}
    
    The power spectra show higher levels of perturbation following the edge of the bubble when projected along the z-axis, but do not when the jet is viewed off-axis, as the bubble clears the inner annuli very quickly when looking at the jet in the x or y-direction. 
    Generally, the amplitude of the PSD decreases with radius from the AGN and with time after the AGN turns off. 
    The spectra measured in the simulation also compare well in shape to the observational spectra.
    
    Given a measured sound speed of $\sim900\,{\rm km/s}$ in the inner cluster and a proportionality coefficient of $\eta$ \textasciitilde 1 \citep{Zhuravleva2014}, the amplitude of density fluctuations are consistent with the \citet{aharonian:18b} turbulence measurement in the cluster from the {\em Hitomi} spectroscopic data at simulation times later than 30 Myr. 
    At later times, as the bubble reaches the outer annuli, the power spectrum extends outward to larger annuli as well, with levels falling below the observations in the outer annuli but comparable in the inner annuli.
    
\begin{figure}
        \centering
            \includegraphics[width=\columnwidth]{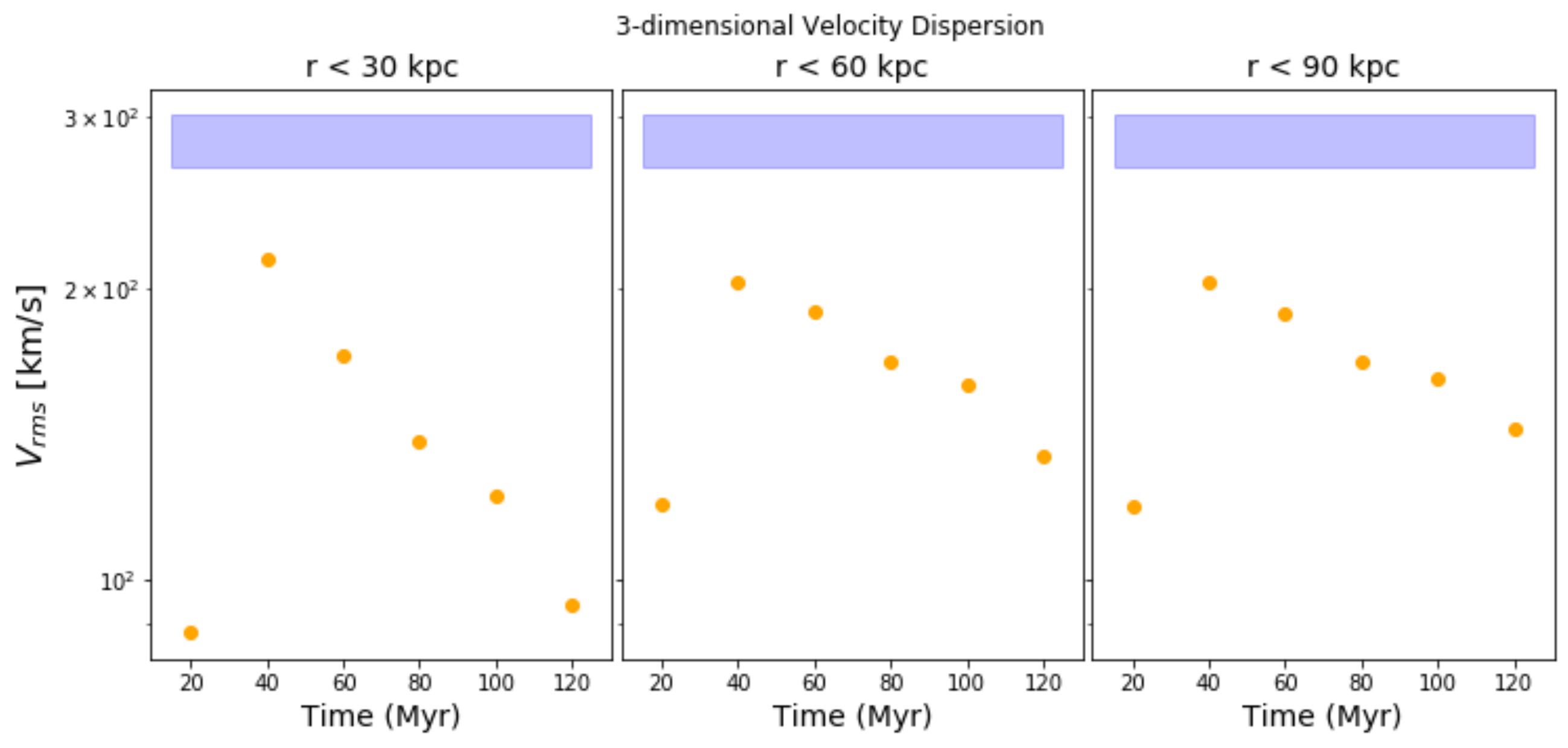}
            \includegraphics[width=\columnwidth]{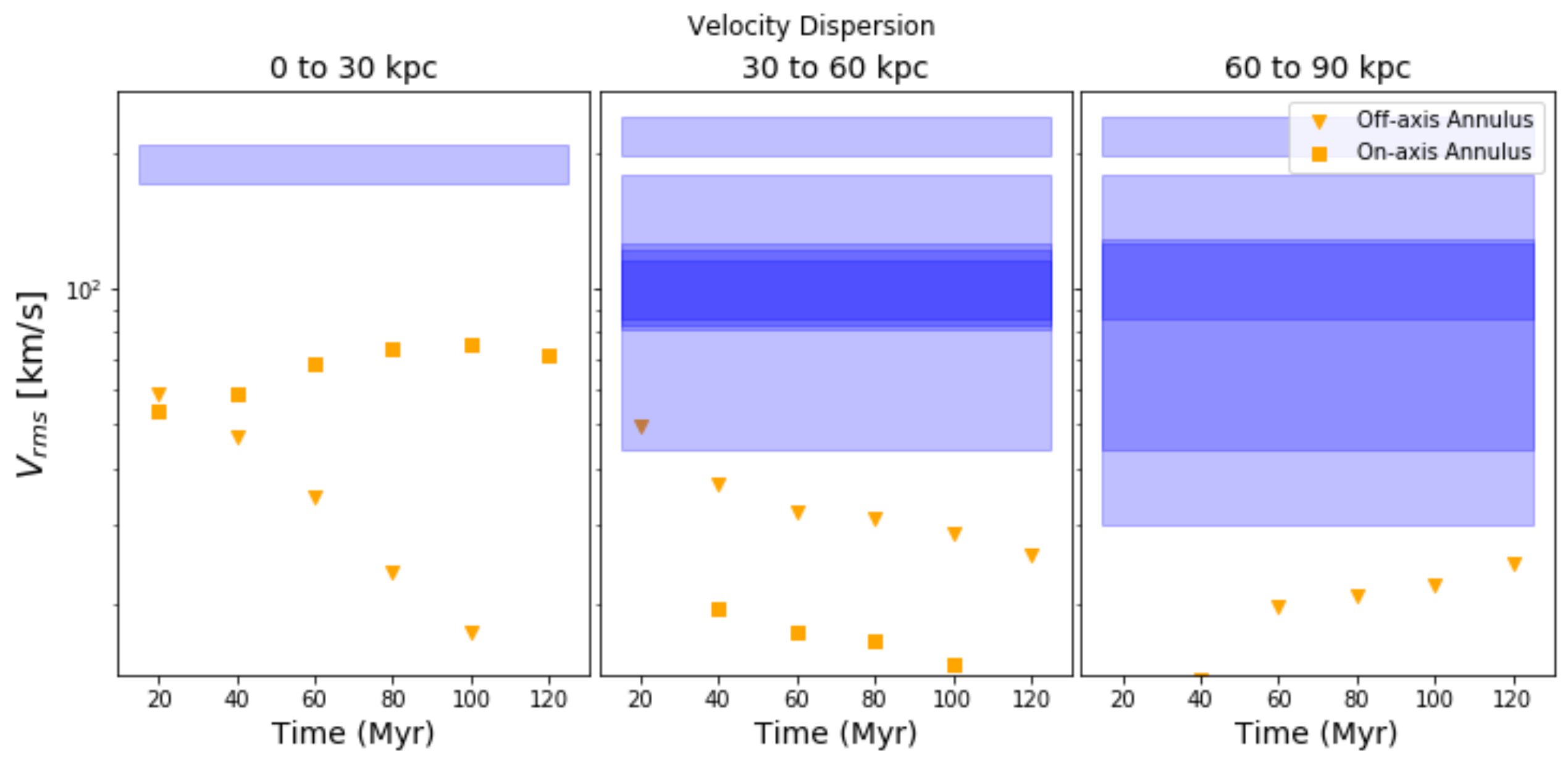}
        \caption{Top: Three-dimensional velocity dispersion of the ICM, weighted by X-ray emissivity, in a sphere around the center of the cluster simulation. The blue-shaded regions correspond to the integrated line-of-sight velocity dispersion at $164\pm10$ km/s from \citet{Aharonian2016}. Bottom: Single-component velocity dispersion, weighted by X-ray emissivity, in an annulus, projected along and perpendicular to the axis of the jet. Blue-shaded regions show the 1-sigma uncertainties of the PSF-corrected line-of-sight velocity dispersions measured by {\em Hitomi} in various regions on the sky overlapping with each respective annulus. The regions are defined and mapped in \citet{aharonian:18b}.}
        \label{fig:veldisp}
    \end{figure}

    To test how our simulated turbulent driving compares to the overall measurements of cluster turbulence, we compared the true measured velocity dispersion from our simulations against the {\em Hitomi} measurement for velocity dispersion in Figure  \ref{fig:veldisp}, finding the RMS velocity extremely variable within the center of the cluster.
    It is approaching the values reported by {\em Hitomi} during and right after the AGN episode, and consistently underpredicting the {\em Hitomi} measurement after \textasciitilde 40 Myr.
    This suggests that in our simulations turbulence is not completely responsible for the observed surface brightness fluctuations in the inner cluster, so that our simple linear assumption of the proportionality constant $\eta=1$ may not hold. 
    Other than turbulence, the impulsive and instantaneous power injection by the jet in our simulation may also lead to stronger shocks and sharper bubble boundaries than might be expected if the inflation were more gradual in the real Perseus cluster, which would appear as additional power in the PSD that would not have corresponding signal in the turbulent velocity dispersion.
    
    \subsection{The Relative Importance of Jets v. Sloshing in Driving Cluster Turbulence}
    \label{sec:sloshing}
    It is important to keep in mind that the analysis we perform fits the first two annuli only. 
    For the outer annuli, the model ensures that the power spectra are not overpredicted. 
    We find that if the MCMC algorithm is tuned to predict the entire spectrum, the inner annuli are fitted at much higher levels than the observations. 
    We choose instead to underpredict the outer annuli and focus the fitting algorithm on the  60-kpc core, where \citet{Walker2018} find the AGN jet must dominate the turbulence profile of the cluster.
    
    The current paradigm in cooling flow modeling is that the AGN is mostly if not wholly responsible for the energy feedback. 
    If heating occurs primarily through dissipation of turbulence stirred up by the AGN, then the power spectra of AGN simulations should roughly match those observed in the cluster. 
    Our analysis demonstrates that this appears to be true within the innermost regions of the cool core of the Perseus cluster, where the measured power spectra match or exceed the observations at most $k$ values, with simulations matching or exceeding the observed PSD and coming reasonably close to matching the observed 3D velocity dispersion.
    
    On the other hand, a single episode of AGN jet activity under the atmospheric conditions of the Perseus cluster cannot readily explain the observed levels of turbulence in the outer regions of the cluster. 
    This complements the result from \citet{Walker2018}, whose simulations of sloshing in the Perseus cluster indicate that most of the observed PSD in the outer cluster can be explained by sloshing alone, without the need for a dominant contribution from the central AGN outside of 60 kpc.
    
    Thus, our simulations support the suggestion by \citet{Walker2018} that the levels of surface brightness fluctuations (as a proxi of turbulence) in the inner 60 kpc of the cluster, can be readily explained by driving from the AGN, though the exact history of AGN jet power cannot be uniquely determined from our work. 
    They are also consistent with the results by \citet{ehlert:20}, which indicate that a single episode of jet activity comparable to the one studied here does not produce a volume-filling level of turbulence as observed by {\em Hitomi}.
    
    Unlike in the case of AGN simulations, the sloshing simulations are based on a single encounter with a sub-cluster that corresponds well with the observations of the outer atmosphere of the Perseus cluster, and as such fitting the spectra to the observed data is unnecessary and a direct comparison can be made to the data. 
    Moreover, while it is clear from GGM-filtered images of Perseus, seen in Figure \ref{fig:ggm}, that the cluster has undergone many episodes of jet activity, sloshing simulations account for all or most of the observed fluctuation power in the outer annuli, and are seen to be visually comparable to the actual {\em Chandra} data \citep[][]{Walker2018}. 
    
    \subsection{Contributions by non-turbulent features to the Power Spectral Density of X-ray Fluctuations}
    It is known that the energy injected into the cluster by the jet in our simulations is not sufficient to balance ICM cooling losses over the entire duration of the simulation. 
    Over the 10 Myr active period, the jet adds $\sim 3*10^{59}$ ergs, which, by itself, could balance the energy losses of the cool core for roughly 20 Myrs \citep{ksanders:07}. 
    To consistently heat the Perseus cluster, ongoing AGN activity comparable to the cooling luminosity would be required (modulated by any efficiency gains by, e.g., jet-mitigated conduction as proposed by \citealt{Chen2019}, or energy leakage due to, e.g., sound waves carrying energy away from the cool core). 
    This is a reasonable compromise, as our simulations do not aim to model a self-consistent feedback loop, but rather to measure the impact of a single episode of jet activity on the cluster as time passes. 
    Other works similarly show that a single-shot injection of energy at the level simulated here is insufficient to raise the level of turbulence to the value observed by {\em Hitomi} throughout the cool core \citep[e.g.][]{ehlert:20}. 
    
    In a complementary work, \citet{hillel:20} perform multi-episode simulations of AGN feedback, modeled as the inflation of cavities by conical outflows, at resolution comparable to our simulations. 
    They argue that jet-driven turbulence alone does not efficiently heat the cluster in their simulations. 
    This is not inconsistent with our results, as other means of energy transfer in AGN feedback are plausible, such as shock heating and the heatpump mechanism discussed in \citep{Chen2019}.

    The energy required to drive the observed level of turbulence observed by {\em Hitomi} exceeds that injected in our simulation by about an order of magnitude, i.e., even if all of the energy injected by the jet were converted to turbulent energy, the single jet episode could explain the observed level of turbulence only within about 50 kpc. 
    Thus, while the RMS velocity measured in the simulation falls within a factor of 2 of the {\em Hitomi} measurement within the inner 60 kpc of the cluster, the simulations significantly under-predict the projected emission-weighted RMS velocity compared to the {\em Hitomi} measurement. 
    This mismatch between the central velocity dispersion (which is of the same order of magnitude as the {\em Hitomi} measurement) and the 1D line-of-sight (LOS) integrated value is not surprising, given that the LOS integrated velocity dispersion includes significant contributions from the outer parts of the cluster that any perturbations by the jet have not reached in our single snapshot and given that at late times, the single-shot injection of energy in our simulation does not account for the likely ongoing level of feedback. 
    Nevertheless, it is important to note that our simulations slightly overpredict the bulk of the surface brightness power spectral density, while slightly underpredicting the central 3D velocity dispersion.
    
    Not surprisingly, this suggests that in our simulations the surface brightness fluctuations in the inner Perseus regions contain a substantial non-turbulent contribution, e.g. from the sharp, coherent intensity edge at the cavity surface, and the X-ray intensity power spectra and the 1D velocity dispersion may not track each other one-to-one.
    Sharp edges associated with the bubbles have to be excised from the data to avoid overestimating the level of turbulence from intensity power spectra.
    The flattening of the surface brightness power spectra could be used as an indicator of significant edge contribution, as illustrated in the analysis of X-ray data in the innermost annulus in Perseus \citep[see Figures 9 and 11 in][]{Zhuravleva2015}.
    
    At later times in the simulation, however, after the bubble wall has passed out of the inner 30 kpc annulus (after approximately 50 Myrs), the power spectra of the off-axis projections (top panels of Figure \ref{fig:annspec}) still contain a substantial amount of power at levels comparable to the observed power spectrum, which we attribute to large scale motions of the thermal gas in the wake of the rising bubble continuing to drive turbulence in the center of the cluster. This supports our conclusion that the surface brightness fluctuations in the cluster center contain contributions from turbulence driven by previous generations of AGN activity, such as those that led to the formation of the ghost cavities detected in Perseus.
    
    Clearly, a simulation with multiple jet episodes or ongoing jet activity would be required to investigate the development of saturated turbulence in the cool core. 
    Such simulations will be numerically challenging at the resolution we impose.
    Our maximum refinement level at the simulation times considered in this analysis corresponds to a resolution element of 120 pc \citep{Chen2019}, well below the resolution scale of {\em Chandra} (and an even higher central resolution of 30 pc during times when the jet is active), while a comparable simulations with ongoing long term jet-driving, performed by \citet{Yang2016}, has a maximum resolution of 1.95 kpc, for example.
    While the higher resolution allows us to generate high-fidelity snapshots of the surface brightness fluctuations and calculate power spectra that are well-resolved, they necessarily limit our ability to investigate the long-term jet activity. 
    Thus our work complements studies like those by \citet{Yang2016} and \citet{lau:17}. 
    
    While \citet{Yang2016} concluded that turbulence drive by AGN jets was insufficient as a source of feedback and below the observed level in Perseus, \citet{lau:17} suggested that long-term injection of AGN jets can drive levels of turbulence comparable to those observed by {\em Hitomi}. 
    Aided by its %significantly 
    higher resolution, our work slots in between, showing that RMS velocities comparable to the {\em Hitomi} result can be excited in the innermost cluster core during and shortly after the jet episode, but to generate volume-filling turbulence throughout the cluster core would require ongoing driving.
    
    \subsection{Caveats}
    While \citet{Walker2018}'s sloshing simulations have a particular time since the interaction that is reliably tied to Perseus, the same is not true of the AGN simulation, as AGN feedback is a continuous process and the detailed injection history of jet power is not well constrained. 
    In order to best compare the simulation to the observed brightness map, we combine multiple jet ages by linearly adding their calculated power spectra. 
    In order to create the multiple-episode facsimile from the jet simulation we must assume that the simulation's perturbations are sufficiently small, such the power spectra add linearly. 
    If that is a good assumption, we can assign a weight to each set of power spectra according to the MCMC fit in order to show what the density fluctuations would look like in a simulation that included the jet firing, shutting off, and firing again, without having to run many simulations with slightly different configurations. 
    This assumption ignores interaction between the different jet bubbles as well as higher-order terms when it comes to combining the power spectra. 
    This should produce viable results assuming AGN duty cycles on the order of 10\%.
    
    One important source of variance when calculating surface brightness power spectra comes from the jet inclination. 
    As is clear from qualitative observations of Perseus (Figure \ref{fig:ggm}), the past episodes of jet activity have all occurred at different jet orientations, and the inclination of the most recent jet in Perseus has proved difficult to determine. 
    It is easily seen that whether the jet is viewed on-axis ("down the barrel") or off-axis has a significant effect on the evolution of the power spectra in Figure \ref{fig:annspec}. 
    
    Finally, it is well-known that resolution effects can limit the fidelity of power spectra derived from turbulence in numerical spectra at wave numbers approaching the resolution scale. 
    However, in our simulations, the inner cluster is always well-resolved compared to the wave numbers probed by the observed and simulated spectra, lending support to the conclusion that the observed spectra are comparable to the simulations and that one or two episodes of jet activity alone can reasonably explain the observed power spectra in the cool core of the cluster. 
    Only the outermost annuli (larger than 120 kpc in radius) are moderately affected by the resolution limit, and only at larger than $k > 0.1 {\rm kpc^{-1}}$, at which the uncertainty of the measured power spectra is too large to provide any constraint.

\section{Conclusions}
    \label{sec:conclusions}
    We perform equivalent analysis on Chandra X-ray observations of Perseus and simulations of a single episode of jet activity from the cluster's AGN jet. 
    By quantifying both via the $\Delta$-variance power spectrum calculation method, we can quantify the surface brightness fluctuation profiles and directly compare simulated AGN-driven turbulent heating to the observed cooling flow gap of the cluster. 
    We fit the characteristic amplitude spectra across the inner 60 kpc of a sample of jet ages, using a markov chain fitting algorithm to match observed power spectra.
    \begin{itemize}
        \item Density fluctuations derived from the synthetic X-ray surface brightness are comparable in magnitude and spectral form to the observed perturbations in the inner 60 kpc core of the intracluster medium from {\em Chandra}. 
        \item Power spectral density values steadily decline as the bubble propagates outward to larger cluster radii.
        \item A linear combination of different aged jets can reproduce the density perturbation spectrum of the cluster core with two or three episodes of jet activity.
        \item These results complement similar analysis of sloshing simulations, leading to the conclusion that while the AGN may be responsible for the intensity fluctuations in the inner 60 kpc of the cluster, the PSD in the outer atmosphere of Perseus is likely dominated by sloshing physics.
        \item We find that in simulations, AGN-driven motions within a 60 kpc sphere approach the values reported by {\em Hitomi} during and right after the AGN episode. 
        However, the line-of-sight-projected 1D velocity dispersion is significantly underpredicted by the numerical model, indicating that a single AGN episode is insufficient to explain the volume-filling level of turbulence, and/or that outside of the 60 kpc sphere, other sources of turbulence likely act in concert with the AGN.
        \item Our simulations roughly match the power spectrum of the intensity fluctuations measured by Chandra in the inner cluster, while under-predicting the 1D LOS velocity dispersion, suggesting that the measured surface brightness fluctuations in our simulations of Perseus are likely driven by a combination of fully established turbulence and non-linear structures, such as shocks and bubbles. 
        This is consistent with \citet{Zhuravleva2015}, who found that removal of the cavities in the central $\sim 30$ kpc annulus of the Perseus image reduced the power spectral density of the X-ray intensity fluctuations up to a factor of 2.
    \end{itemize}

\section*{Acknowledgements}

    We would like to acknowledge helpful discussions with Marsha Wolf, Eric Hooper, and Eric Wilcots. 
    SH and AH would like to acknowledge support from NASA through the Astrophysics Theory Program grant NNX17AJ98G and National Science Foundation through grant AST1616101. 
    This work used the Extreme Science and Engineering Discovery Environment (XSEDE) Stampede at the Texas Advanced Computing Center at The University of Texas at Austin and the HPC Cluster at the Center for High Throughput Computing at the University of Wisconsin-Madison. 
    Support for this research was provided by the Hilldale Undergraduate Research Fellowship and the Wisconsin Space Grant Consortium.

%%%%%%%%%%%%%%%%%%%%%%%%%%%%%%%%%%%%%%%%%%%%%%%%%%
\section*{Data Availability}

 %   The inclusion of a Data Availability Statement is a requirement for articles published in MNRAS. Data Availability Statements provide a standardised format for readers to understand the availability of data underlying the research results described in the article. The statement may refer to original data generated in the course of the study or to third-party data analysed in the article. The statement should describe and provide means of access, where possible, by linking to the data or providing the required accession numbers for the relevant databases or DOIs.

Simulated images and averaged simulation data products used here will be shared on reasonable request to the corresponding author.



%\bibliographystyle{mnras}
%\bibliography{specpaper}

%\begin{thebibliography}{99}
%\end{thebibliography}

%%%%%%%%%%%%%%%%%%%%%%%%%%%%%%%%%%%%%%%%%%%%%%%%%%

%%%%%%%%%%%%%%%%% APPENDICES %%%%%%%%%%%%%%%%%%%%%

%%%%%%%%%%%%%%%%%%%%%%%%%%%%%%%%%%%%%%%%%%%%%%%%%%

% Don't change these lines
\bsp	% typesetting comment
\label{lastpage}
\end{document}